\documentclass[aps,showpacs,superscriptaddress,twocolumn]{revtex4}

\usepackage{bbm}
\usepackage{amsfonts}
\usepackage{amsmath}
\usepackage{amsthm}
\usepackage{amscd}
\usepackage{amssymb}
\usepackage{amsxtra}

\begin{document}

\title{Key rate for calibration robust entanglement based BB84 quantum key distribution protocol}

\pacs{03.67.Mn, 03.65.Ud, 03.67.Dd}
\keywords{Quantum key distribution, entanglement verification}

\author{O.~Gittsovich}
\affiliation{
Institute for Quantum Computing, University of Waterloo, Waterloo, Ontario, N2L 3G1, Canada
}
\author{T.~Moroder}
\affiliation{
Naturwissenschaftlich-Technische Fakult\"at, Universit\"at Siegen, Walter-Flex-Str. 3, D-57068 Siegen, Germany
}

\begin{abstract}
We apply the approach of verifying entanglement, which is based on the sole knowledge of the
dimension of the underlying physical system to the entanglement based version
of the BB84 quantum key distribution protocol. We show that the familiar one-way
key rate formula holds already if one assumes the assumption that
one of the parties is measuring a qubit and no further assumptions about the measurement are needed.
\end{abstract}

\maketitle

\section{Introduction}
Entanglement verification is known to be a prerequisite
sub-protocol in quantum key distribution (QKD) \cite{marcos}.
In physical implementations of QKD protocols,
it is often hard to justify a particular theoretical assumption;
no matter how advanced our current technologies are, one inevitably faces
experimental imperfections. For example, in the entanglement based version
of the BB84 protocol \cite{bennett84a,shorpreskill}, one assumes that the parties are able to perform
projective measurements in two mutually unbiased bases.
This cannot be realized with an absolute certainty in
experiments. Abandoning this assumption leads to a completely device independent
version of QKD, where the only way to prove security is to show that
correlations violate one of the Bell inequalities.
However, this case can appear as another extreme, where the complete knowledge of
the system is substituted by complete ignorance of what the system might be.

A certain degree of trust in the underlying physical system is, however, not so
far off from the reality. The example given is a cold atom in an ion trap. This
system can be considered as a qubit with a very high precision.
In Ref. \cite{us}, we developed an approach for entanglement verification based
on different forms of partial information, like knowing the underlying dimension or
further measurement properties such as sharpness or orthogonality; for other alternatives
see Ref. \cite{rosset11pra}. In this note we apply
this method in order to derive a key formula for the entanglement based
BB84 protocol. In particular, we show that if one registers the standard observations
(symmetric bit error and no correlations if one measures in different bases) then
one can use the same one-way key rate formula as in the completely characterized scheme.

\section{Key rate formula}
In the BB84 setting, Alice and Bob perform two different dichotomic $\pm$ measurements each, which give rise to
random variables (data) ${\bf X}$, ${\bf X}'$ and ${\bf Y}$, ${\bf Y}'$ at Alice's and Bob's stations respectively.
For the description of a hypothetical situation, when Alice is able to perform sharp and orthogonal measurements, i.e.,
standard projections in two mutually unbiased bases, we use random variables $\overline{\bf X}$, $\overline{\bf X}'$.
Here $\overline{\bf X}$ results from the perfect projective measurement that gave rise to the random variable
${\bf X}$ \cite{us}, while $\overline{\bf X}'$ results from any projective measurement
performed in a mutually unbiased bases with respect to the
one that produces $\overline{\bf X}$. This means that ${\bf X}$ can be achieved
from $\overline{\bf X}$ by performing a classical postprocessing.
The key rate using one-way classical postprocessing from Alice to Bob is given by \cite{devetak05a}
\begin{align}
R_{\rightarrow}&\geq \min[I({\bf X}:{\bf Y})-I({\bf X}:{\bf E})]\nonumber\\
&\geq I({\bf X}:{\bf Y})-\max I(\overline{\bf X}:{\bf E})\nonumber\\
&\geq I({\bf X}:{\bf Y}) - \max(1-H(\overline{\bf X}|{\bf E}))\nonumber\\
&\geq I({\bf X}:{\bf Y}) - \max H(\overline{\bf X}'|{\bf Y}),
\label{eq:keyrate}
\end{align}
where the first inequality is the data processing inequality $I({\bf X}:{\bf E})\leq I(\overline{\bf X}:{\bf E})$,
the second readily follows from the definition of
the mutual information $I$ and from the bound $H({\bf X})\leq -\log|{\bf X}|=1$ and the third
is the entropic uncertainty relation \cite{tomamichel11}. Note that at this stage one does not need
the quantum description of the measurements on Bob's side anymore.

The data which Alice and Bob record after the quantum
stage of the protocol can be presented in a data matrix of the form
\begin{equation}
D=\left(\begin{array}{ccc}
1 & \mathbb{E}({\bf Y}) & \mathbb{E}({\bf Y}')\\
\mathbb{E}({\bf X}) & \mathbb{E}({\bf XY}) & \mathbb{E}({\bf XY}')\\
\mathbb{E}({\bf X}') & \mathbb{E}({\bf X}'{\bf Y}) & \mathbb{E}({\bf X}'{\bf Y}')
\end{array}\right),
\end{equation}
which completely characterizes the probability distribution.
This matrix can be transformed to another data matrix which contains the hypothetical
random variables $\overline{\bf X}$ and $\overline{\bf X}'$ on Alice's side by
$\overline{D}=R\cdot S\cdot D$ \cite{us}, with
\begin{equation}
S=\left(\begin{array}{ccc}
1 & 0 & 0\\
x_1 & x_2 & 0\\
x_3 & 0 & x_4
\end{array}\right),\;
R=\left(\begin{array}{ccc}
1 & 0 & 0\\
0 & 1 & 0\\
0 & -\cot\theta & \csc\theta
\end{array}\right),
\end{equation}
with real parameters $x_2\geq1+|x_1|$, $x_4\geq1+|x_3|$ that characterize unsharpness of Alice's measurements and
$\theta\in[0,\pi]$ the relative angle between the measurements' directions.

Assume now that we observe the data which is common in the BB84 protocol, i.e., symmetric bit error, no
correlations if one measures in different bases and uniform marginals. This corresponds
to a diagonal matrix
\begin{equation}
D=\left(\begin{array}{ccc}
1 & 0 & 0\\
0 & \sigma & 0\\
0 & 0 & \sigma
\end{array}\right)
\end{equation}
and hence one obtains
\begin{equation}
\overline{D}=\left(\begin{array}{ccc}
1 & 0 & 0\\
x_1 & \sigma x_2 & 0\\
x_3 & -\sigma x_4 \cot\theta & \sigma x_4\csc\theta
\end{array}\right).
\end{equation}
First of all, we note that $H({\bf X}|{\bf Y})=h_2(\frac{1-\sigma}{2})$ with
$h_2(x)=-x\log_2x-(1-x)\log_2(1-x)$. Second,
from $\overline{D}$ we can extract the upper bound on $\max H(\overline{\bf X}'|{\bf Y}')$
\begin{align}
\max H(\overline{\bf X}'|{\bf Y}')&\leq\max h_2\left[\frac{1}{2}(1-\mathbb{E}(\overline{\bf X}'{\bf Y}'))\right]\nonumber\\
&\leq\max_{x_4,\theta}\left[h_2\left(\frac{1}{2}(1-\sigma x_4 \csc\theta)\right)\right]\label{eq:estimate}\nonumber\\
&=h_2\left(\frac{1-\sigma}{2}\right),
\end{align}
where the first inequality follows from the classical data processing inequality, while
the last equality follows from the minimization with constraints $x_4\geq1,\sin\theta\in[0,1]$. By noting
that $Q=(1-\sigma)/2$ corresponds to quantum bit error rate after parameter estimation and plugging Eq. (\ref{eq:estimate})
into Eq. (\ref{eq:keyrate}), we arrive at the same key rate formula as in Ref. \cite{shorpreskill}:
\begin{equation}
R_{\rightarrow}\geq 1 - 2 h_2\left(Q\right).
\end{equation}

\section{Acknowledgements}
We thank N. L\"utkenhaus and O. G\"uhne for the discussions.
This work has been supported by the EU (Marie Curie CIG
293993/ENFOQI) and the BMBF (Chist-Era Project
QUASAR). Oleg Gittsovich is especially grateful for the support of
the Austrian Science Fund (FWF) and Marie Curie Actions
(Erwin Schr\"odinger Stipendium J3312-N27).

\bibliographystyle{aipproc}   

\end{document}